\documentclass[10pt,twocolumn]{article}
\usepackage{graphicx}

\begin{document}

\title{Systematic errors and combination of individual CRF solutions
in the framework of the international pilot project for the next ICRF}
\author{Julia Sokolova, Zinovy Malkin \\ Central (Pulkovo) Astronomical Observatory RAS, \\
Pulkovskoe~Ch. 65-1, St.~Petersburg 196140, Russia}
\date{March 07, 2007}
\maketitle

\begin{abstract}
A new international Pilot Project for the re-determination of the ICRF was
initiated by the International VLBI Service for Geodesy and Astrometry (IVS)
in January 2005. The purpose of this project is to compare the individual CRF
solutions and to analyze their systematic and random errors with focus on the
selection of the optimal strategy for the next ICRF realization. Eight radio
source catalogues provided by the IVS Analysis Centers GA, SHAO, DGFI,
GIUB-BKG, JPL, MAO NANU, GSFC, USNO were analyzed.
In present study, four analytical models were used to investigate the
systematic differences between solutions: solid rotation, rotation and
deformation (IERS method), and expansion in orthogonal functions:
Legendre-Fourier
polynomials and spherical functions. It was found that expansions by
orthogonal function describe the differences between individual catalogues
better than the two former models.
Finally, the combined CRF was generated. Using the radio source positions from
this combined catalogue for estimation of EOP has shown improvement of the
uncertainty of the celestial pole offset time series.

\end{abstract}


\section{Introduction}
\label{intro}

Celestial reference system (CRF) as realized by a set of
coordinates of selected celestial objects is widely used for
numerous astronomy, navigation, time and other measurements. The
CRF accuracy and stability are all-important for successful
solution of all these tasks. For millennia, the CRF was based on
optical observations and star positions. With establishment of new
observing technique, very long baseline interferometry (VLBI),
much more accurate CRF realization became available.
In 1998, the CRF based on the positions of extragalactic radio sources
has been adopted by IAU (International Astronomy Union)
as the fundamental celestial reference frame, replacing the
FK5 optical frame (Arias et~al. 1995, Ma et~al. 1998).

After publishing of the first VLBI RSC, attempts was made to
improve the accuracy of radio-band CRF by means of constructing of
combined catalogues, as it was customary for optical astronomy,
where fundamental catalogues served as an international standard
for astrometry and other measurements on the sky.  Different
methods were used to obtain a combined RSC, e.g. Walter (1989a,b),
Yatskiv \& Kur'yanova (1990), Kur'yanova \& Yatskiv (1993). Also,
till 1995, IERS (International Earth Rotation Service) used
derived combined RSC for maintenance of the IERS Celestial
Reference Frame.

However, starting from 1996, new CRF realization was adopted by
the IERS, and further approved by the IAU in 1998.  The first
realization of the ICRF was based on the refined analysis of VLBI
observations made at the NASA Goddard Space Flight Center, USA (Ma
et al. 1998).  All the 608 radio sources included in the ICRF was
divided into three groups: 212 {\it defining} sources whose
coordinates are supposed to be kept in the future realizations to
maintain the ICRF orientation, 294 {\it candidate} sources not
sufficiently monitored, and 102 {\bf other} sources for improving
the sky coverage.  In 1999 and 2004, two ICRF extensions ICRF-Ext.1
(Ma 2001) and ICRF-Ext.2 (Fey et al. 2004) have been published.
In those versions, the positions of 212 {\it defining} sources
were kept the same as obtained in the first ICRF. It should be
noted that both ICRF extensions were obtained in a manner similar
to the first realization, {\it i.e.} as a result of analysis of
the VLBI observations at a single analysis center.  The latest
ICRF realization, ICRF-Ext.2, is hereafter referred to as ICRF.

In the end of 2004, joint pilot project of the IERS and the IVS
(International VLBI Service for Geodesy and Astrometry, Schlueter
et al. 2002) was initiated (Ma 2004, Call for Participation). One
of the main goals of the project was to seek after possible ways
to improve the existing ICRF. Large experience accrued by the
optical astrometry during centuries shows that combining
catalogues of the star positions have better random and systematic
accuracy than individual catalogues. In particular, the latter can
be affected by the systematic errors caused by algorithms and
software used for data processing. Hopefully, a combining
procedure can be used to mitigate influence of errors of
individual RSC. Another main goal of the Pilot Project is to
develop new methods of comparison of RSC adequate to the modern
level of their precision and accuracy, in other words, their
random (stochastic) and systematic errors.

In this paper, the work was made in four steps.
\begin{enumerate}
\item Analysis of the random and systematic errors of individual (input)
   catalogues, and a choice of the most adequate method of representation
   of the systematic differences between catalogues.
\item Determination of the systematic differences between the input catalogues
   and ICRF.
\item Construction of a combined catalogue in the ICRF system (stochastic
   improvement of the ICRF).
\item Construction of the final combined catalogue (systematic improvement of
   the ICRF).
\end{enumerate}

First, we searched after optimal method of representation of the
systematic errors of the RSC. Then we investigated a possibility of improving
the ICRF by means of combining individual RSC.
Four methods of representation of the systematic part of differences
in the RSC have been examined on the basis of comparison of the residual
differences in the radio source coordinates. Eight individual radio source
catalogues, and ICRF were used in this study.
After the most accurate method had been chosen, it was used to compute
the systematic differences between the individual catalogues and ICRF.
Finally, these
differences were used in the procedure of construction of combined RSC.
Thus obtained combined catalogue was tested by means of computation
of celestial pole offset time series with both combined and ICRF RSC.
Result of this test showed improvement of the scatter of the time series
when combined RSC is used.


\section{Input catalogues}
\label{input}

Input catalogues used in this study were submitted by eight IVS Analysis
Centers:
AUS (Geoscience Australia), BKG (Bundesamt f\"ur Kartographie und Geod\"asie,
Germany), DGFI (Deutsches Geod\"atisches Forschungsinstitut, Germany),
GSFC (NASA Goddard Space Flight Center, USA), JPL (Caltech/NASA Jet Propulsion
Laboratory, USA), MAO (Main Astronomical Observatory of National Academy
of Sciences of Ukraine), SHAO (Shanghai Astronomical Observatory, China),
USNO (U.~S.~Naval Observatory, USA).
Brief description of the input catalogues is given in
Table~\ref{tab:input_cat}.

\begin{table*}
\centering
\caption{Input catalogues. The last column shows number of
the sources in the catalogue and number of reference sources
used to tie the orientation of the catalogue to ICRF.}
\label{tab:input_cat}
\begin{tabular}{llccl}
\hline\hline
Center & Software & Time span  & \multicolumn{2}{c}{Number of} \\
       &          & month/year & delays & sources   \\
\hline
AUS  & OCCAM       & 11/1979 -- 12/2004 & 3208197 & 737(207) \\
BKG  & Calc/Solve  & 01/1980 -- 01/2005 & 4031453 & 748(212) \\
DGFI & OCCAM       & 01/1980 -- 01/2005 & 3650771 & 686(199) \\
GSFC & Calc/Solve  & 08/1979 -- 01/2005 & 4574189 & 954(212) \\
JPL  & MODEST      & 10/1978 -- 01/2005 & 3575847 & 734(2) \\
MAO  & SteelBreeze & 10/1980 -- 01/2005 & 3773765 & 685(25) \\
SHAO & Calc/Solve  & 04/1980 -- 01/2005 & 4431503 & 813(212) \\
USNO & Calc/Solve  & 09/1979 -- 01/2005 & 4252684 & 943(207) \\
\hline
\end{tabular}
\end{table*}

\begin{figure}
\centering
\includegraphics[height=\hsize,viewport=260 350 400 750]{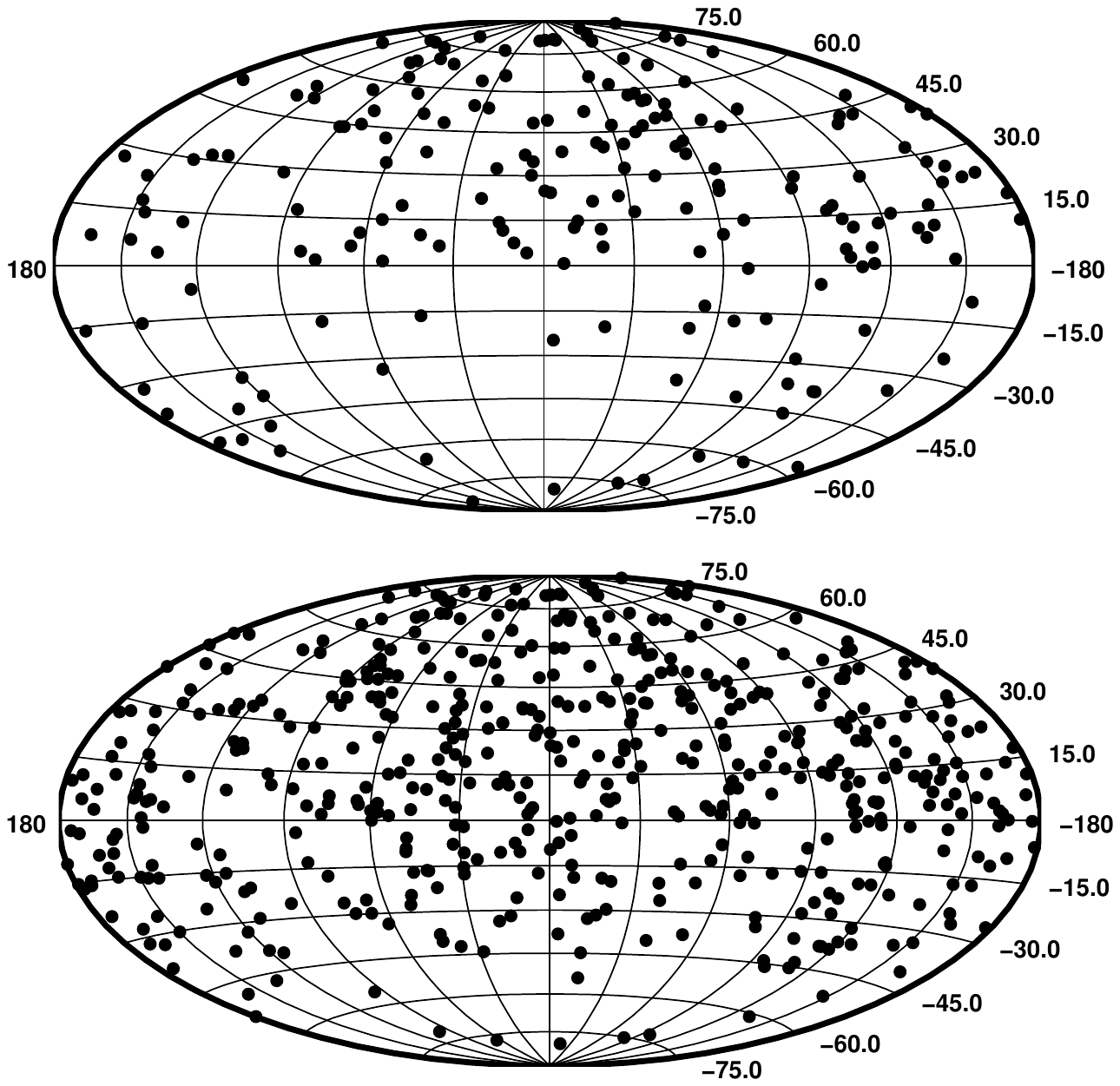}
\caption{Sky distribution for 196 {\it defining} ({\it top}) and
all 525 sources ({\it bottom}) common for the input catalogues.}
\label{fig:sky_distr}
\end{figure}

Usually, investigation of the systematic differences between catalogues is made
using some set of reference sources common for compared catalogues.
Comparison of the lists of radio sources included in the input catalogues
showed that there are 525 sources present in all the catalogues, 196 of them
belong to the list of 212 ICRF {\it defining} sources.
This should be mentioned that we took into account only the sources which
have at least 15 observations in 2 sessions.  After such a selection,
the total number of sources present in all the input catalogues amounts to 968.
Fig.~\ref{fig:sky_distr} shows the distribution of common sources over the sky.

Both source lists, as well as other subsets of 525 sources, can be
used as reference for analysis of the systematic differences
between catalogues. All the computations described below were
carried out for both 196 and 525 sources. At this paper, we present
only the results obtained with the first list of 196 sources.
Although definite differences in results were found, the main
conclusions made in this study do not depend on the reference source list.

\begin{figure}
\centering
\includegraphics[width=\hsize,viewport=150 560 510 740]{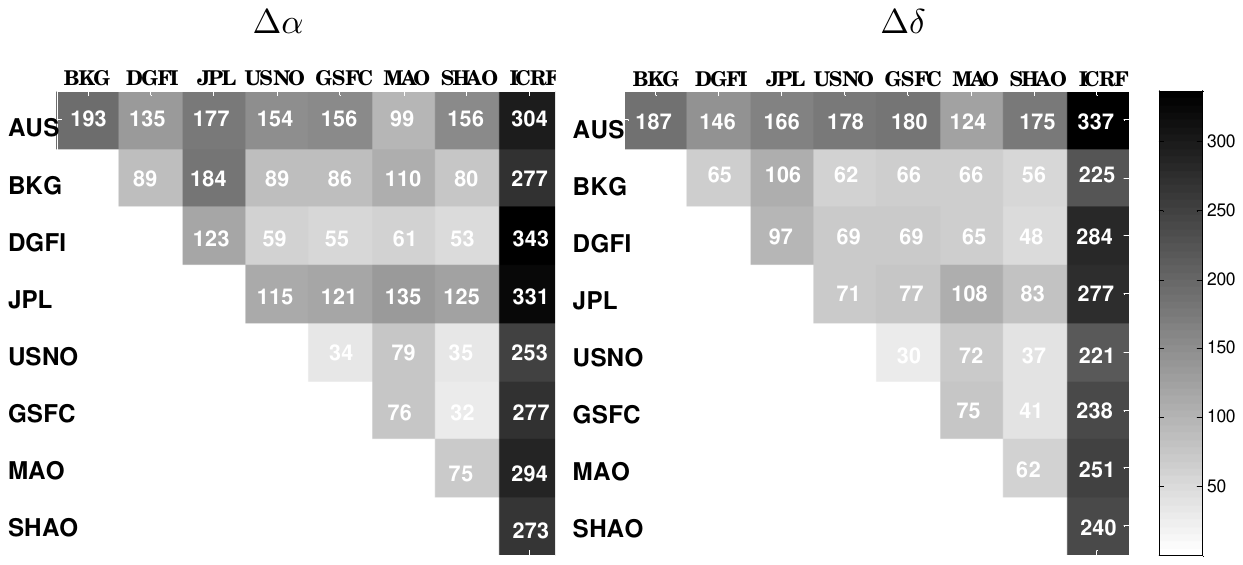}
\caption{WRMS differences between the input catalogues and ICRF. Unit: $\mu$as.}
\label{fig:wrms_raw}
\end{figure}

Weighted root-mean-square (WRMS) differences of the radio source
coordinates between the input catalogues and ICRF are shown in
Fig.~\ref{fig:wrms_raw} ($\mu$as stands for microarcseconds).
One can see from Fig.~\ref{fig:wrms_raw} WRMS
differences have the least values for catalogues computed with
Calc/Solve software, both for intercomparison of these catalogues
and their comparison with ICRF. The latter most probably is caused
by the fact that the ICRF was constructed using Calc/Solve
software. Large WRMS differences between JPL and other catalogues
may be caused by its orientation to ICRF which has been defined by only
two reference sources, unlike other catalogues, for which much
longer lists of reference sources were used. Catalogue AUS shows the
greatest differences with other catalogues, probably, because it
is the only catalogue constructed using the Least Squares
Collocation method, while other Analysis Centers used conventional
Least Squares. One can see that the DGFI catalogue, also constructed
using OCCAM software, but Least Squares version, does not stand
out against other catalogues.

The WRMS differences between the input catalogues themselves may be
not valuable, especially when they are caused by the difference in
orientation of the catalogues' axes, which can be easily accounted for
during combination. More interesting fact is that all the input
catalogues demonstrate rather large differences with the ICRF, which
may indicate significant systematic errors in the latter. This
preliminary conclusion will be confirmed by the analysis of the
systematic differences given below.


\section{Analytical representation of the systematic differences}
\label{sysdif}

Suppose, we have two catalogues of position of celestial sources given as set
of spherical
coordinates $\alpha_{1i},\delta_{1i}$ and $\alpha_{2i},\delta_{2i}$,
where $i$ is the source number.
Denote the differences between spherical coordinates of the $i$-th object given in
two catalogues as
$\Delta\alpha_i=\alpha_{1i}-\alpha_{2i}, \Delta\delta_i=\delta_{1i}-\delta_{2i}$.
The aim of an analytical representation of the systematic differences between
two catalogues is to obtain formulae

\begin{equation}
\begin{array}{ll}
&\Delta\alpha = f_\alpha\,(\alpha,\delta) \,, \\
&\Delta\delta = f_\delta\,(\alpha,\delta) \,, \\[1ex]
&{\rm or} \\[1ex]
&\left\{\begin{array}{c}\Delta\alpha \\ \Delta\delta \end{array} \right\} =
f\,(\alpha,\delta) \,,
\end{array}
\label{eq:common}
\end{equation}
\noindent which provides minimum residuals between $\Delta\alpha_i, \, \Delta\delta_i$
observed and computed analytically.
In fact, such a representation of the systematic differences is a kind of
a low-pass filter, which allow us to smooth stochastic errors in coordinates.
Having such a representation, one can reduce a RSC to the system of another
catalogue.

In this section, we compare four methods of analytical representation of the
systematic differences between radio source positions given in various
catalogues.  Those methods are: simple rotation around three Cartesian axes
hereafter referenced to as ``R'', rotation plus deformation used by IERS
(''RD``), Brosche's method (``B'') and expansion in Legendre-Fourier
functions (``LF'').

It should be noted that in usual astrometric practice differences between
source positions in right ascension are used as $\Delta\alpha\cos\delta$,
which reflects the geometry of the celestial sphere.  However, hereafter
we use $\Delta\alpha$ because IERS's method is formulated only for this type
of differences.  Other method listed below can be easily adapted to
$\Delta\alpha\cos\delta$.

\subsection{Simple rotation}

Two given catalogues realize two Cartesian coordinate systems
$X_1Y_1Z_1$ and $X_2Y_2Z_2$.
Then differences between source positions in two catalogues
can be represented as the result of rotation of the second coordinate system with
respect to the first coordinate system about axes $XYZ$ by three angles $A_1, A_2, A_3$.
Then the systematic differences between two catalogues can be expressed as
(Walter \& Sovers, 2000)
\begin{equation}
\begin{array}{rl}
\Delta\alpha \!\!\!&=\! \phantom{-}A_1\tan\delta\cos\alpha + A_2\tan\delta\sin\alpha - A_3 \,, \\[1ex]
\Delta\delta \!\!\!&=\! -A_1\sin\alpha + A_2\cos\alpha \,.
\end{array}
\label{eq:rotation}
\end{equation}

\subsection{Rotation with deformation}

This method of representation of the systematic differences between
radio source catalogues was proposed by Arias and Bouquillon (2004),
and it is used by the IERS since 1995.  The authors added to
(\ref{eq:rotation}) three supplement terms to account for some
specific errors of VLBI catalogues. In this method, the systematic
differences between two catalogues are approximated by
\begin{equation}
\begin{array}{rl}
\Delta\alpha \!\!\!&= \! \phantom{-}A_1\tan\delta\cos\alpha + A_2\tan\delta\sin\alpha - A_3  + D_{\alpha}\delta \,,\\
\Delta\delta \!\!\!&= \! -A_1\sin\alpha + A_2\cos\alpha + D_{\delta} + B_{\delta} \,.
\end{array}
\label{eq:iers}
\end{equation}

\subsection{Brosche's method}

In this method, as well as in the next one, an expansion of the
differences in source positions between catalogues in orthogonal
functions is used. Large experience collected by the optical
astrometry proved that such an expansion provides the highest
accuracy of the representation of the systematic errors of
catalogues of celestial source positions. In this case,
the general representation of the differences (\ref{eq:common})
is given by
\begin{equation}
\left\{\begin{array}{c}\Delta\alpha \\ \Delta\delta \end{array} \right\} =
\sum_{j=0}^{g}b_jY_j(\alpha,\delta) \,,
\label{eq:ortho}
\end{equation}
\noindent where $b_j$ are the coefficient to be found from analysis of the differences.
According to Brosche (1966)
\begin{equation}
Y_j(\alpha,\delta) = \left\{\begin{array}{ll}
P_{n0}(\delta), & k=0, l\ne1 \,, \\[1ex]
P_{nk}(\delta)\sin(k\alpha), & k \ne 0, l=0 \,, \\[1ex]
P_{nk}(\delta)\cos(k\alpha), & k \ne 0, l=1 \,,
\end{array}
\right.
\label{eq:brosche}
\end{equation}
\noindent where $P_{nk}$, associated Legendre polynomials are given by
\begin{equation}
P_{nk}(\delta)=\cos^k\delta\left[\sin^p\delta+\sum_{\mu=1}^{[p/2]}
\frac{(-1)^\mu\prod\limits_{\nu=0}^{2\mu-1}(p-\nu)}{\prod\limits_{\nu=1}^{\mu} 2\nu(2n-2\nu+1)}
\sin^{p-2\mu}\delta\right].
\label{eq:legendre}
\end{equation}
\noindent where $p=n-k$, \ $[p/2]$ is entier of $p/2$, \ $n=0,1,\ldots$,
\ $k=0,1,2,\ldots,n$, \ $j=n^2+2k+l-1$.

\subsection{Legendre-Fourier functions}

Bien et al. (1978) has proposed to use another set of orthogonal functions for
better representation of systematic differences between catalogues, especially
in the polar regions. In this case (\ref{eq:ortho}) is given by
\begin{equation}
\begin{array}{rcl}
\left\{\begin{array}{c}\Delta\alpha \\ \Delta\delta \end{array} \right\}& = &
\sum\limits_{nkl}b_{nkl}Y_{nkl}(\alpha,\delta) \,, \\[1em]
Y_{nkl}(\alpha,\delta) & = & R_{nkl}L_n(\sin\delta)F_{kl}(\alpha) \,.
\end{array}
\label{eq:lf1}
\end{equation}

Here we omit the Hermite function included in the original expression of Bien et al.
(1978) to account for the source brightness.
Commonly speaking, such a dependence may do exist in the case of VLBI
observations too, and it worth separate investigating.
Legendre polynomials can be computed using recursion
\begin{equation}
\begin{array}{l}
L_0=1 \,, \\[1ex]
L_1=\sin\delta \,, \\[1ex]
L_{n+1}(\sin\delta)=
  \frac{\displaystyle 2n+1}{\displaystyle n+1}\sin\delta\, L_n(\sin\delta)-
  \frac{\displaystyle n}{\displaystyle n+1}L_{n-1}(\sin{\delta})\,,\\[1ex]
\qquad n=2,3,\ldots \,.
\end{array}
\label{eq:lf2}
\end{equation}

Fourier functions are given by
\begin{equation}
F_{kl}(\alpha)=\left\{
\begin{array}{lll}
1,& k=0, & l=-1 \,, \\[1ex]
\cos(kl\alpha),  & k \ne 0, & l=1 \,, \\[1ex]
\sin(-kl\alpha), & k \ne 0, & l=-1 \,.
\end{array}
\right.
\label{eq:lf3}
\end{equation}

Lastly, normalizing functions are given by
\begin{equation}
R_{nkl}=\sqrt{2n+1}
\left\{
\begin{array}{ll}
1, & k=0 \,,\\
\sqrt{2}, & k \ne 0.\\
\end{array}
\right.
\label{eq:lf4}
\end{equation}

\begin{figure}
\centering
\includegraphics[width=\hsize,viewport=150 230 350 710]{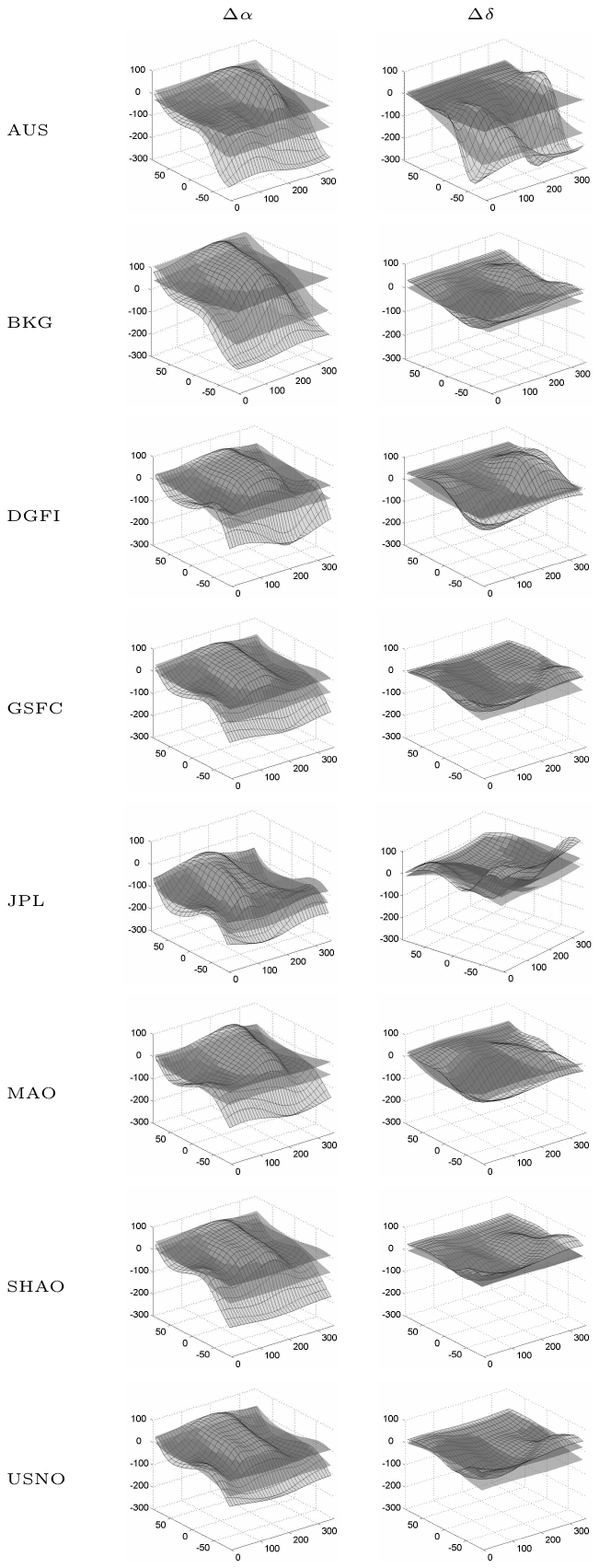}
\caption{Analytical representation of the differences between the
input catalogues and ICRF:
R (dark grey), RD (grey), LF (light grey).
Original differences are shown in black lines.
Horizontal axes show right ascension ({\it right}) and
declination ({\it left}) in degrees. Unit: $\mu$as.}
\label{fig:comp_icrf}
\end{figure}

All the four methods described above were applied to the differences
between each of eight input catalogues and ICRF for 196 common
{\it defining} sources (see section~ref{input}.
For this purpose, the coefficients of (\ref{eq:rotation}), (\ref{eq:iers}),
(\ref{eq:brosche}), (\ref{eq:lf1}) were found by means of Least Squares
adjustment.
After that we computed the residuals between original differences
and those computed by formulae (\ref{eq:rotation}), (\ref{eq:iers}),
(\ref{eq:brosche}), (\ref{eq:lf1}).
The results are presented in Fig.~\ref{fig:comp_icrf}
and Table~\ref{tab:comp_wrms}.
Representation with the Brosche's model is not shown
in Fig.~\ref{fig:comp_icrf} because corresponding surface
practically coincides with the Legendre-Fourier expansion.
One can see that expansion
in Legendre-Fourier functions provides the least residuals, {it i.e.} most
accurate representation of the systematic differences between catalogues.
Expansion in spherical functions (Brosche's method) gives worse
accuracy.  As to the first two methods, they seem to be not adequate
to actual errors of modern RSC.

\begin{table*}
\centering
\caption{WRMS residuals between the input catalogues and ICRF before (Raw)
and after approximation of the systematic differences (see notation of the
methods in text). Results related to the LF method providing
the best approximation are shown in bold. Unit: $\mu$as.}
\label{tab:comp_wrms}
\begin{tabular}{lcccccccc}
\hline\hline
& AUS & BKG & DGFI & JPL & USNO & GSFC & MAO & SHAO \\
\hline
&\multicolumn{8}{c}{$\Delta\alpha$} \\
Raw     &     304 &     277 &     343 &     331 &     253 &     277 &     294 &     273 \\
R       &     301 &     271 &     342 &     308 &     249 &     274 &     286 &     271 \\
RD      &     299 &     265 &     342 &     308 &     247 &     273 &     285 &     270 \\
B       &     170 &     177 &     237 &     238 &     172 &     191 &     203 &     193 \\
{\bf LF}&{\bf 106}&{\bf 125}&{\bf 164}&{\bf 172}&{\bf 122}&{\bf 144}&{\bf 152}&{\bf 145}\\
&\multicolumn{8}{c}{$\Delta\delta$} \\
Raw     &     337 &     225 &     284 &     277 &     221 &     238 &     251 &     240 \\
R       &     337 &     225 &     284 &     273 &     221 &     238 &     251 &     240 \\
RD      &     333 &     224 &     283 &     273 &     221 &     237 &     251 &     239 \\
B       &     180 &     159 &     178 &     182 &     152 &     158 &     169 &     166 \\
{\bf LF}&{\bf 111}&{\bf 109}&{\bf 112}&{\bf 127}&{\bf 104}&{\bf 106}&{\bf 134}&{\bf 111}\\
\hline
\end{tabular}
\end{table*}


\section{Combined catalogue in the ICRF system}
\label{impr_random}

\begin{figure}
\centering
\includegraphics[width=\hsize,viewport=150 560 510 720]{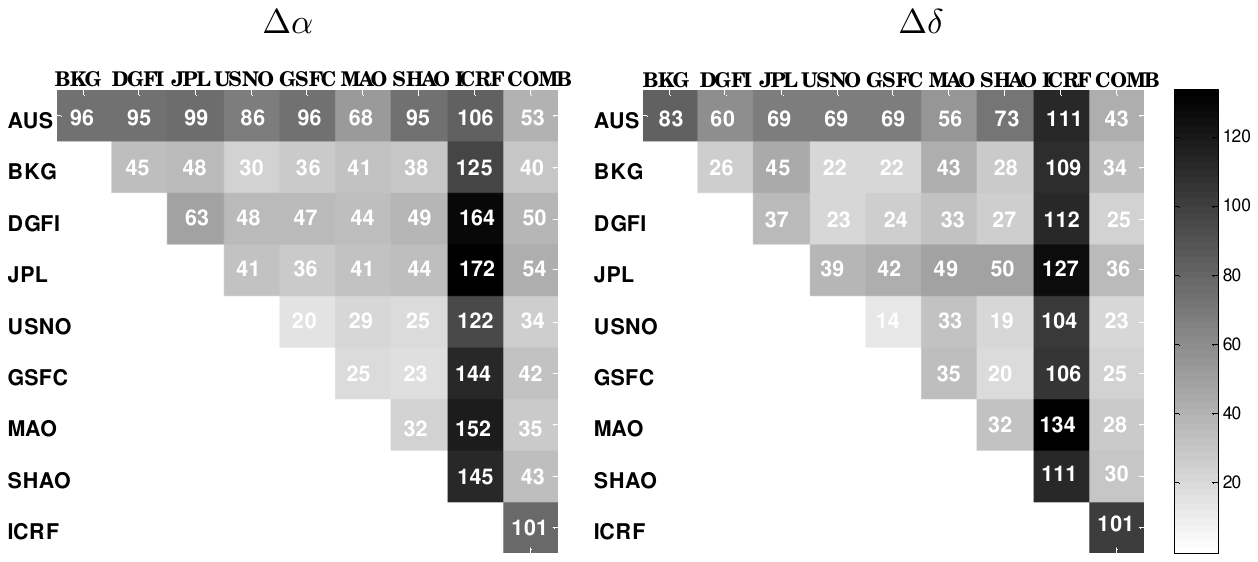}
\caption{WRMS differences between the input catalogues transformed to the ICRF
system and combined catalogue RSC(PUL)07C01. Unit: $\mu$as.}
\label{fig:wrms_comb1}
\end{figure}

At the next step, the systematic differences between the input
catalogues and ICRF found by the LF method as described in the
previous section were applied to all of the input catalogues with
the view to transforming them to the ICRF system. After that,
the coordinates of all the sources in transformed catalogues were
averaged with weights depending on the formal errors of coordinates.
In result, the combined catalogue RSC(PUL)07C01 was constructed.
This catalogue containing all the 968 sources present in the input
catalogues can be considered as a stochastic improvement of the
ICRF. Fig.~\ref{fig:wrms_comb1} shows the WRMS differences between
the input catalogues transformed to the ICRF system and
RSC(PUL)07C01.

Comparing Fig.~\ref{fig:wrms_raw} and \ref{fig:wrms_comb1}, one can
conclude that the differences between the input catalogues contain
not only systematic part described by analytical representation,
but also, in some cases, significant stochastic and/or
high-frequency components.

\begin{figure}
\centering
\includegraphics[width=\hsize,viewport=160 600 440 720]{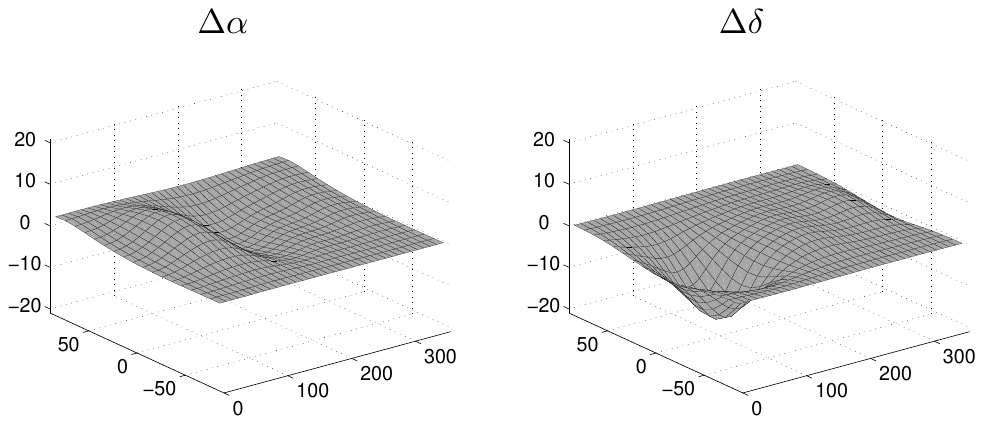}
\caption{Differences between RSC(PUL)07C01 and ICRF.  Horizontal
axes show right ascension ({\it right}) and declination ({\it
left}) in degrees. Unit: $\mu$as.} \label{fig:comb1-icrf}
\end{figure}

Fig.~\ref{fig:comb1-icrf} shows the systematic difference between
combined catalogue RSC(PUL)07C01 and ICRF.  One can see that the
catalogue RSC(PUL)07C01 represents the ICRF system at a level of
about 10 $\mu$as.

\section{Final combined catalogue}
\label{impr_system}

Final combined catalogue was constructed in the following way.
Let us call the system of a catalogue the set of coefficients
of Legendre-Fourier functions obtained for given catalogue.
Thus we have eight systems corresponding to the eight input catalogues.
At the first iteration, these systems were averaged without weights.
Then the WRMS differences between this average system and the systems
of the input catalogues for the bins
$10^{\circ}(\alpha)\times 5^{\circ}(\delta)$
were computed. Thus obtained WRMS were used for weighting of the input
catalogues at the second iteration.
Although we use different weights of the input catalogues for
different bins, they are close for each catalogue.
Final weights of the catalogues
averaged over the sky are given in Table~\ref{tab:weights}.

\begin{table*}
\centering
\caption{Weights of the input catalogues applied during combination,
averaged over the sky.}
\label{tab:weights}
\begin{tabular}{lcccccccc}
\hline\hline
         & AUS   & BKG   & DGFI  & GSFC  & JPL   & MAO   & SHAO  & USNO \\
\hline
$\alpha$ & 0.246 & 0.671 & 0.464 & 1.993 & 0.254 & 0.558 & 1.792 & 2.062 \\
$\delta$ & 0.205 & 1.220 & 0.586 & 1.921 & 0.446 & 0.541 & 1.459 & 1.927 \\
\hline
\end{tabular}
\end{table*}

\begin{figure}
\centering
\includegraphics[width=\hsize,viewport=150 560 510 720]{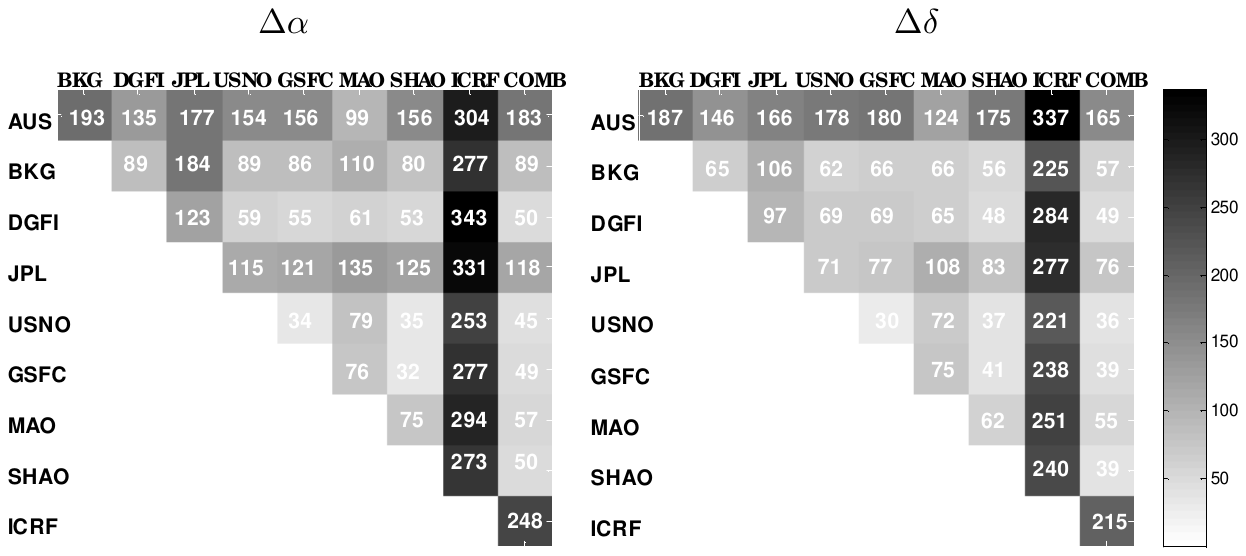}
\caption{WRMS differences between the input catalogues
and combined catalogue RSC(PUL)07C02. Unit: $\mu$as.}
\label{fig:wrms_comb2}
\end{figure}

\begin{figure}
\centering
\includegraphics[width=\hsize,viewport=150 230 350 710]{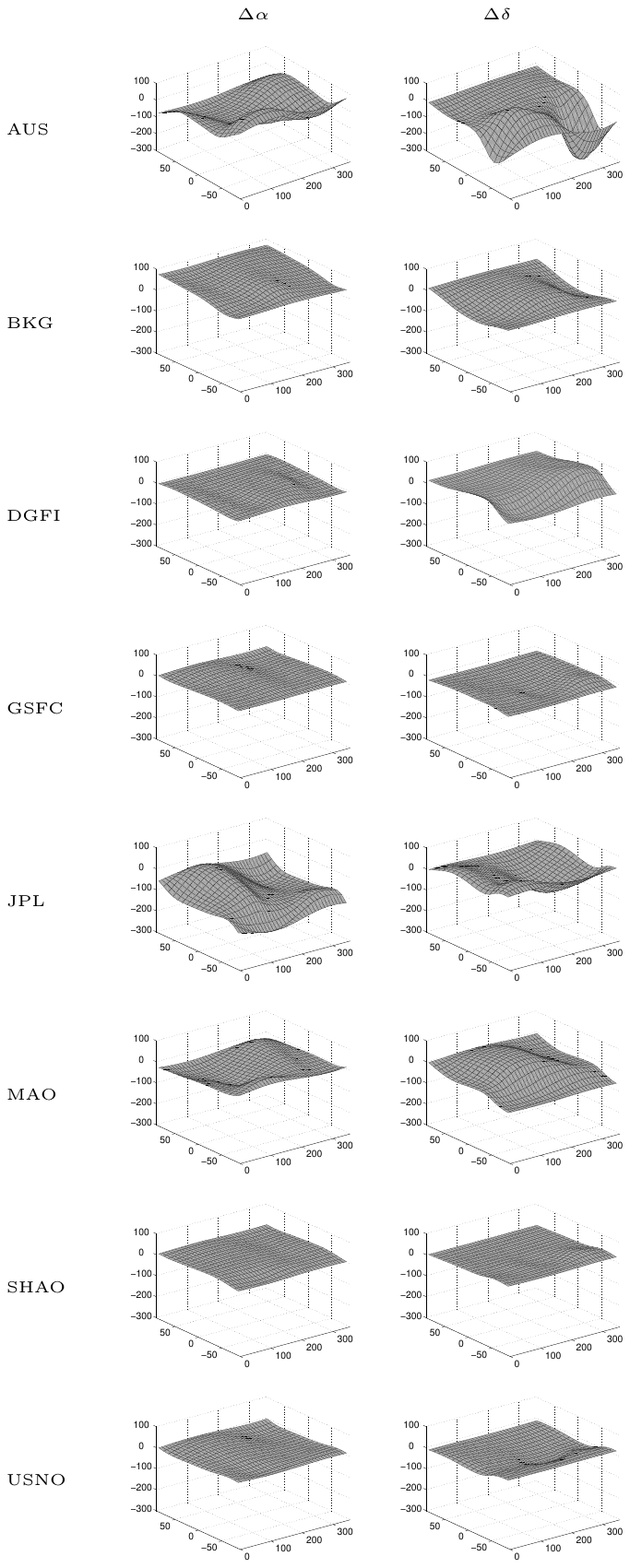}
\caption{Systematic differences between the input catalogues and
combined catalogue RSC(PUL)07C02.  Horizontal axes show right
ascension ({\it right}) and declination ({\it left}) in degrees.
Unit: $\mu$as.} \label{fig:diff_comb2}
\end{figure}

Thus computed average system was added to the first combined catalogue
RSC(PUL)07C01.
In result, final combined catalogue, RSC(PUL)07C02 have been obtained.
It can be considered as both stochastic and systematic improvement
of the ICRF. In Fig.~\ref{fig:wrms_comb2}, WRMS differences between
the input catalogues and combined catalogue RSC(PUL)07C02 are shown,
and Fig.~\ref{fig:diff_comb2} presents the systematic differences
between the input catalogues and RSC(PUL)07C02.

\begin{figure}
\centering
\includegraphics[width=\hsize,viewport=150 600 450 730]{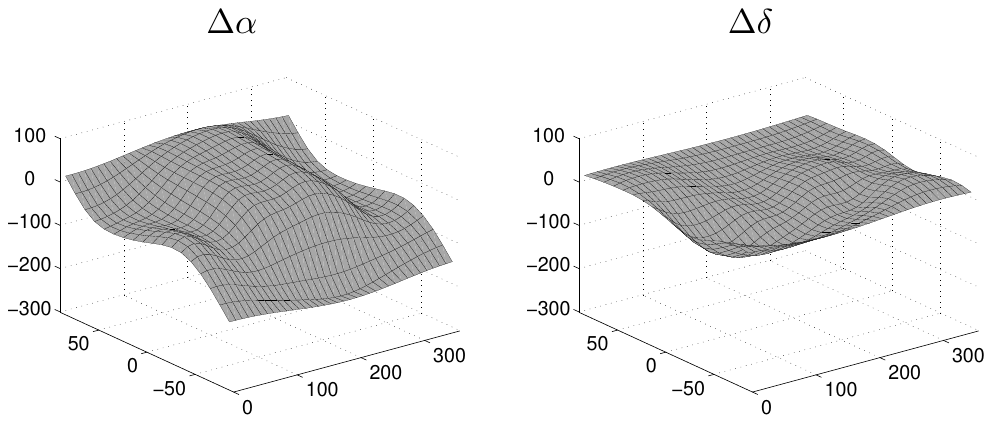}
\caption{Differences between RSC(PUL)07C02 and ICRF.  Horizontal
axes show right ascension ({\it right}) and declination ({\it
left}) in degrees. Unit: $\mu$as.} \label{fig:comb2-icrf}
\end{figure}

Comparison of RSC(PUL)07C02 and ICRF is presented in Fig.~\ref{fig:comb2-icrf},
which show the result of expansion of the differences between two catalogues
in Legendre-Fourier functions. Results of this comparison lead us to the
supposition that ICRF may have significant systematic errors.

\section{Comparison with observations}
\label{comp_obs}

It is important to assess an actual accuracy of obtained catalogue
(as well as other CRF realizations).
Unfortunately, existing methods of comparison of catalogues allow us
to investigate only {\it differences} between catalogues.
Here, we use a test, which can help us to get some independent estimate
of the quality of our combined catalogue.
For this purpose, we compute two celestial pole offset time series
from processing of R1 and R4 IVS observing programs observed in
the period 2002--2006 with two radio
source catalogues, ICRF-Ext.2 and RSC(PUL)07C02.
Then we compute the WRMS differences between computed celestial pole
offsets and Free Core Nutation model.
The result of this test presented in Table~\ref{fig:scatter}
shows clear improvement of the scatter of celestial pole offset estimates.

\begin{table}
\centering
\caption{WRMS differences between celestial pole offset series computed with
two CRF realization and IAU2000A model corrected for FCN contribution.
Unit: $\mu$as.}
\label{fig:scatter}
\begin{tabular}{lccc}
\hline\hline
Catalogue & $X_c$ & $Y_c$ \\
\hline
ICRF-Ext.2    & 103 & 100 \\
RSC(PUL)07C02 & ~98 & ~98 \\
\hline
\end{tabular}
\end{table}


\section{Conclusion}
\label{concl}

In this paper, we have constructed a new combined catalogue of
radio source coordinates. For this study we used eight catalogues
submitted by IVS Analysis Centers in the framework of the IERS/IVS
Pilot Project on the future realization of ICRF.

First, we have examined four methods of analytical representation
of systematic differences between catalogues of radio source
coordinates. Representation by means of expansion in
Legendre-Fourier functions is proved to be the most accurate
method. Two methods usually used for comparison of radio source
catalogues, simple rotation and rotation with deformation seem to
be not suitable for investigation of modern radio source
catalogues, and should be replaced by more adequate one.

Two combined radio source catalogues have been constructed.  The
first of them is obtained as weighted average of the input
catalogues corrected for systematic differences with ICRF. It can
be considered as stochastic improvement of the current realization
of ICRF. Second combined catalogue have been obtained from the
first one after applying the weighted average systematic
difference between the input catalogues and ICRF, which allows one
to account also for possible ICRF systematic errors.

To compare our combined catalogue with ICRF, we used two tests,
which allow us to independently estimate the scatter of celestial
pole offsets time series obtained from processing of VLBI
observations using ICRF and combined catalogue. Both tests have
showed improvement of the results.

The results obtained in this paper allow us to make a conclusion
that ICRF-Ext.2 may have significant systematic errors, most probably caused
by fixing the coordinates of 212 {\it defining} sources to its values obtained
in the first ICRF version of 1995.

Further development of this study may include:
\begin{itemize}
\item More detailed analysis of stochastic and systematic errors
of radio source catalogues.
\item Estimation of possible impact of the high-frequency systematic errors
in source position on the orientation of the catalogue axes.
\item Analysis of the reasons of systematic differences between
radio source catalogues.
\item Careful selection of the input catalogues and reference sources.
\item Elaboration of weighting method.
\item Development of more sensitive methods of the assessment of the accuracy
of CRF realizations.
\end{itemize}

It is also interesting to compare different methods of
construction of combined catalogue of radio source coordinates, in
particular a method used so-called arc approach developed by
Yatskiv \& Kur'yanova (1990).

\section{Acknowledgements}
The authors greatly appreciate all the IVS Analysis Centers providing
their catalogues of the position of radio sources used in this study.

\end{document}